\documentclass[10pt,letterpaper]{article}
\usepackage{opex3,cite}
\begin{document}

\title{Transmission degradation and preservation for tapered optical fibers in rubidium vapor}
\author{M.M. Lai, J.D. Franson, and T.B. Pittman}
\address{Physics Department, University of Maryland Baltimore County, Baltimore, MD 21250 USA}
\email{meimei.lai@umbc.edu}

\begin{abstract} The use of sub-wavelength diameter tapered optical fibers (TOF's) in warm rubidium vapor has recently been identified as a promising system for realizing ultra-low-power nonlinear optical effects. However, at the relatively high atomic densities needed for many of these experiments, rubidium atoms accumulating on the TOF surface can cause a significant loss of overall transmission through the fiber. Here we report direct measurements of the time-scale associated with this transmission degradation for various rubidium density conditions. Transmission is affected almost immediately after the introduction of rubidium vapor into the system, and declines rapidly as the density is increased. More significantly, we show how a heating element designed to raise the TOF temperature can be used to reduce this transmission loss and dramatically extend the effective TOF transmission lifetime.
\end{abstract}



\section{Introduction}
\label{sec:introduction}

The desire for nonlinear optics using ultra-low-power fields is motivated by applications ranging from nanophotonic switching \cite{almeida04} to quantum computing with single-photon qubits \cite{milburn89}. One promising system for realizing these effects is the use of sub-wavelength diameter Tapered Optical Fibers (TOF's) suspended in Rubidium vapor.  These systems have recently been used to demonstrate nW-level saturated absorption and EIT \cite{spillane08}, two-photon absorption \cite{hendrickson10}, and all-optical modulation \cite{salit11}. Related work includes the use of hollow-core photonic band-gap fibers filled with Rb vapor \cite{ghosh06,bajcsy09,londero09,saha11,venkataraman11}, and various integrated waveguides in contact with Rb vapor \cite{yang07,wu10,hendrickson12,stern12}. In all of these systems, low-power nonlinearities are enabled by the high intensities achieved through compression of the optical mode-area interacting with the Rb atoms.

One of the major technical issues in these systems is the accumulation of Rb on the waveguide surfaces \cite{stern12,slepkov08,hendrickson09}.  This can be particularly problematic for the case of TOF’s, where surface contaminants are known to cause scattering of the guided evanescent mode and, consequently, a degradation of transmission through the fiber \cite{fujiwara12}. Because high transmission is essential for low-power nonlinear optics applications, the loss due to Rb accumulation is currently one of the main limiting factors in these kinds of experiments. In order to move towards more practical systems, a better understanding of transmission degradation for TOF's in Rb vapor is important at the present time.

In this paper, we describe an experiment which allowed real-time observations of TOF transmission degradation as the density of the surrounding Rb vapor was increased. We also show that a heating element used to raise the surface temperature of the TOF higher than the surrounding environment can reduce the accumulation of Rb atoms, thereby reducing the amount of transmission loss. Our results indicate that it is possible to achieve relatively high TOF transmission, even in the presence of the relatively high Rb vapor densities needed for many low-power nonlinear optics applications. This systematic study of the potential effectiveness of TOF heating was motivated by the use of basic heating elements in some of the original papers \cite{spillane08,hendrickson10}.

\begin{figure}[t]
\centering\includegraphics[width=4in]{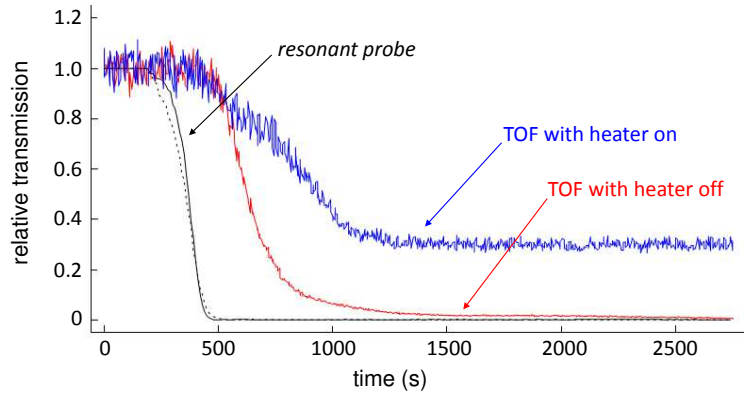}
\caption{A comparison of two runs measuring transmission degradation for TOF's as Rb vapor is rapidly introduced into the system. The solid black line shows the absorption of an auxiliary free-space resonant probe beam; a substantial increase in Rb density occurs roughly 300 seconds into the run. The red curve shows the transmission of a TOF with the heating unit turned off; a significant loss of TOF transmission occurs with a time-scale on the order of several minutes. A second run done in virtually identical conditions (blue curve) used a TOF with the heating element turned on (the dashed black line shows the resonant probe for this run). The TOF transmission degradation is seen to be much less severe with the heater turned on. The vertical axis is normalized so the average transmission values are 1 before the introduction of Rb vapor.}
\label{fig:mainresults}
\end{figure}

The main results of our study are summarized in Fig. \ref{fig:mainresults}. Sub-wavelength diameter TOF's were mounted on a custom heating unit, and installed in a vacuum system. Transmission through the TOF's was then monitored as Rb vapor was  introduced into the system. The rapidly increasing Rb density was monitored using the absorption of an auxiliary free-space resonant probe beam. The red curve in Fig. \ref{fig:mainresults} shows the results of a run with the TOF heating unit turned off. A rapid drop in overall TOF transmission is seen within several minutes after the introduction of significant Rb density. In contrast, the blue curve shows the results of a virtually identical run with the TOF heating unit turned on. In this case the transmission drop is significantly smaller, leaving the TOF in a much more usable condition for the experiments of interest.  The details of these TOF ``heater off'' vs. ``heater on'' measurements will be discussed in the following sections.

The remainder of the paper is organized as follows: in Section \ref{sec:setup} we describe the design of the TOF heating unit and the overall experimental setup, and confirm the system's suitability for low-power nonlinear optics applications by demonstrating Rb saturation with nW-level fields \cite{spillane08}. In Section \ref{sec:results} we present detailed measurements of transmission degradation and preservation for three different scenarios, and in Section \ref{sec:TOFfailures} we briefly describe suspected TOF failure mechanisms that prevented indefinite use of the system.

\section{TOF heating unit and experimental setup}
\label{sec:setup}

Our TOF's were produced from standard single-mode fiber using the well-known ``flame-brush'' technique with an air-propane flame \cite{birks92}. The TOF's typically had a minimum diameter of roughly 300 nm, and a sub-500 nm diameter over a length of about 5 mm. The overall transmission of freshly produced TOF's was typically in the range of 40 - 70\%.

\begin{figure}[t]
\centering\includegraphics[width=4.25in]{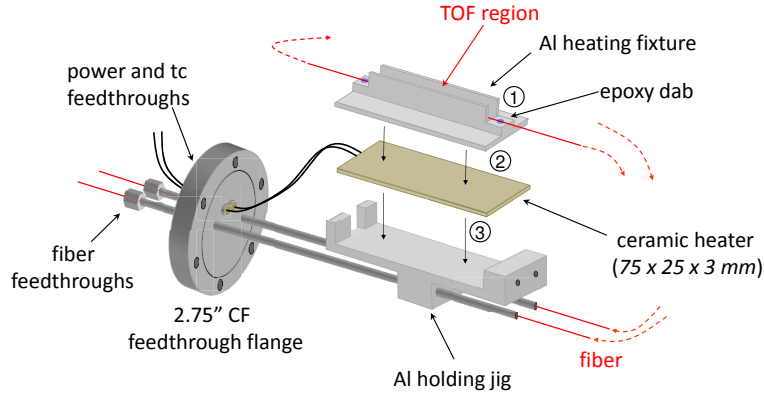}
\caption{Overview of the TOF heating unit used to elevate the surface temperature of the TOF. Installation of the TOF involved 3 main steps: in Step \textcircled{1}, a freshly prepared TOF is mounted in an aluminum ``canyon shaped'' heating fixture using UV curable epoxy. In Step \textcircled{2}, the heating fixture is placed on top of a ceramic heating element, and in Step \textcircled{3} both are pressed tightly into an aluminum holding jig and secured using set-screws. The holding jig is mounted on a 2.75$''$ ConFlat (CF) feedthrough flange that allows the fiber to pass into and out of the vacuum system \protect\cite{abraham98}. Power and thermocouple (tc) feedthroughs are used to control the TOF heating.}
\label{fig:heater}
\end{figure}

Figure \ref{fig:heater} shows an overview of the TOF heating unit used in our experiments.  The TOF's were mounted in an aluminum ``canyon shaped'' heating fixture using small dabs of low-outgassing UV curable epoxy. The TOF region was suspended above the canyon floor, and radiatively heated by the walls and floor on three sides; Rb vapor entered the canyon from the open top side. Conduction heating through the nanofiber itself also contributed to the elevated TOF surface temperature.

The aluminum heating fixture was placed on top of a $75 \times 25$ mm ceramic heater with an integrated thermocouple (Watlow  CER-1-01-00007), and held in an aluminum holding jig that was mounted on the stainless steel tubes of a 2.75$''$ ConFlat (CF) feedthrough flange with exterior Swagelok fittings. These Swagelok fittings used Teflon feedthroughs for coupling the fiber in and out of the flange \cite{abraham98}. The flange also had power and thermocouple feedthroughs for controlling the ceramic heater.

\begin{figure}[t]
\centering\includegraphics[width=4.5in]{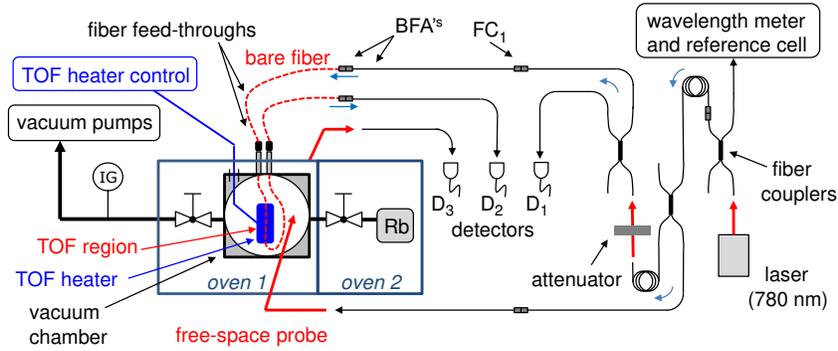}
\caption{Overview of the experimental setup. In the optical part of the setup, the thin black lines represent single-mode fiber patch cords and couplers, while the thicker red arrows denote free-space beams. The TOF was installed inside the vacuum chamber with the external bare-fiber leads connected to the optical setup using two bare-fiber adapters (BFA's). One of the standard fiber patch cord connectors (labelled FC$_{1}$) is identified due to its importance in the measurements. Additional details can be found in the main text.}
\label{fig:setup}
\end{figure}

An overview of the vacuum system is shown on the left side of Fig. \ref{fig:setup}. The feedthrough flange was installed on the top side of a standard 4.5$''$ CF six-way cube that served as the main vacuum chamber. The base pressure (measured by an Ion gauge near the pumps) was typically $\sim 10^{-7}$ Torr after baking out the system. Two large ovens were used to independently control the temperature of the main chamber, and a metallic Rb source (Alpha-Aesar 10315). In order to rapidly introduce Rb vapor into the chamber, we developed a repeatable procedure where the chamber was held at $\sim$ 60$^{o}$C, and the Rb source oven temperature was quickly increased to a value of $\sim$ 200$^{o}$C at the beginning of a run. In our system, this particular heating sequence caused a rapid increase in Rb density followed by fairly steady-state conditions lasting longer than the duration of the data runs of interest.

As shown in the lower right side of Fig. \ref{fig:setup}, the optical part of the system was driven by a tunable external cavity diode laser at 780 nm (New Focus Velocity 6312). A series of three fiber couplers split the laser into four signals: one to a wavelength meter and Rb reference cell, a second to the TOF, a third was used as a laser power reference (detector $D_{1}$), and the fourth was used as the free-space probe beam that passed through windows on the sides of the vacuum chamber. The TOF and free-space probe transmissions were monitored using detectors $D_{2}$ and $D_{3}$, respectively. A variable attenuator between two of the fiber couplers allowed us to reduce and control the TOF input powers to the desired nW power-levels.

\begin{figure}[t]
\centering\includegraphics[width=5in]{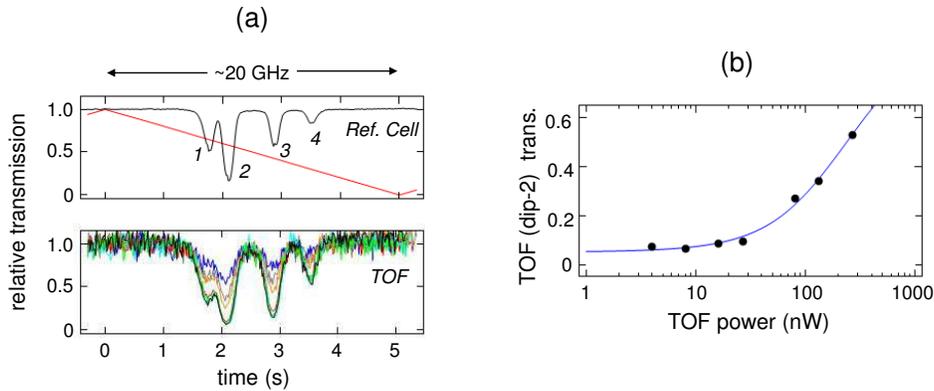}
\caption{Spectroscopy and ultra-low power saturation using TOF's in Rb vapor \protect\cite{spillane08}. The upper panel in part (a) shows the absorption spectrum (with dips labelled 1-4) of the Rb D$_{2}$ line at 780 nm obtained in a standard reference cell, while the lower panel shows the corresponding absorption spectrum in the TOF signal for seven different TOF power levels (different color traces). Part (b) shows the TOF ``dip 2'' transmission values as a function of power; the data is fit by a simple nonlinear absorption model with a saturation power of only 72 nW. The main point of this data is to confirm that we have the kind of TOF in Rb vapor system needed for low-power nonlinear optics experiments \protect\cite{spillane08,hendrickson10,salit11}. }
\label{fig:Psat}
\end{figure}

Using this setup we were able to simultaneously perform Rb absorption spectroscopy using three different systems: (1) the auxiliary reference cell signal, (2) the free-space probe beam signal, and (3) the TOF signal.  The laser frequency was slowly swept (5 seconds) over a 20 GHz range centered on the Rb D$_{2}$ absorption line at 780 nm.  Figure \ref{fig:Psat}(a) shows the familiar absorption dips resulting from the four ground state hyperfine levels of $^{85}$Rb and $^{87}$Rb  \cite{steck12}. The upper panel shows the absorption spectra obtained with the auxiliary reference cell, while the lower panel shows spectra obtained with the TOF for seven different power levels ranging from 4 - 267 nW.

The decreasing depth of the TOF dips with increasing power shows the ability to saturate the Rb vapor with extremely low powers, as first shown in the pioneering paper by Spillane et.al. \cite{spillane08}.  Figure \ref{fig:Psat}(b) shows a Log-Linear plot of the ``dip 2'' ($^{85}$Rb, $F_{g}=3$) transmission data vs. estimated power in the TOF. The data is fit (blue line) by a simple transmission model $T = e^{-\alpha_{NL} L}$, with a nonlinear absorption coefficient defined as $\alpha_{NL} = \alpha/(1+P/P_{sat})$, and a best-fit saturation power of $P_{sat} = $ 72 nW.  The ability to saturate the Rb vapor at such low powers arises from the reduced optical mode areas in the TOF waist region (on the order of 1 um$^{2}$ \cite{tong04}), and agrees with calculated saturation powers (on the order of 10 - 100 nW \cite{spillane08}) for these kinds TOF in Rb vapor systems. 

We note that our ability to accurately determine the TOF powers interacting with the Rb vapor was hindered somewhat by non-ideal transmission \cite{frawley12} of the TOF's themselves (67\% in the present example), and simple fiber connector losses in the optical system. To avoid repeatedly disconnecting the fragile bare-fiber adapters (BFA's, Fig. \ref{fig:setup}), we measured the overall system transmission by comparing the input powers at the more robust connector FC$_{1}$ with the output powers at detector $D_{2}$. The TOF powers were then estimated by dividing the output power by an average BFA connector throughput value of 75\% (determined by auxiliary tests).

\section{Transmission degradation and preservation measurements}
\label{sec:results}

\begin{figure}[t]
\centering\includegraphics[width=5in]{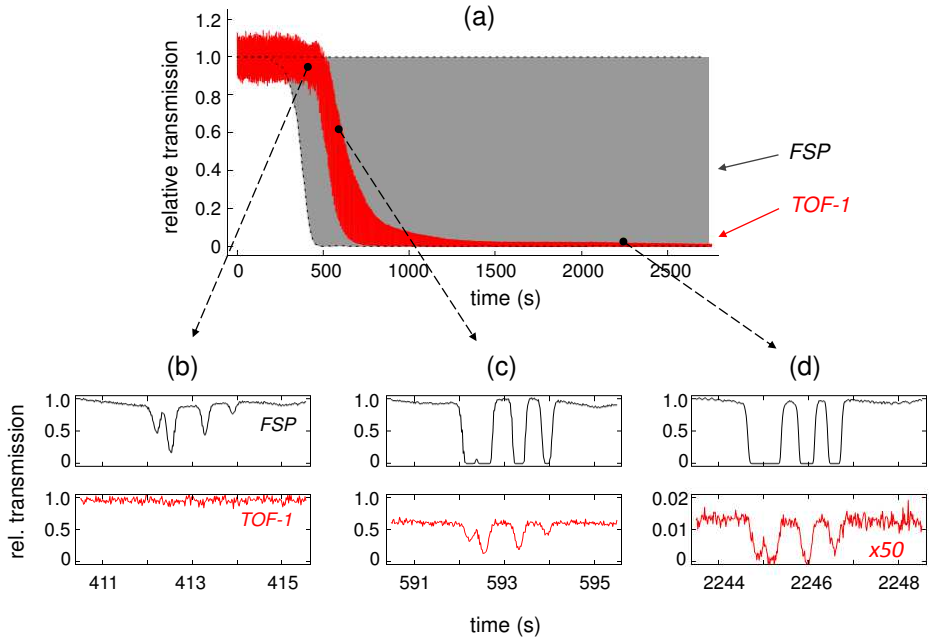}
\caption{Detailed results of a run with the TOF heater unit turned off. The laser is swept back and forth across the Rb D$_{2}$ line as Rb vapor is introduced into the system. In part (a), the 2700 second run represents 540 consecutive five-second sweeps through the Rb absorption dips; the grey trace is the free-space probe beam (FSP) signal, and the red trace is the TOF (TOF-1) signal. Parts (b)-(d) zoom-in on individual scans through the Rb absorption lines at three points during the run. The TOF-1 transmission degrades to 1.5\% of its initial value by the end of the run, but still shows strong Rb absorption dips (part (d)).}
\label{fig:TOF1}
\end{figure}

\begin{figure}[t]
\centering\includegraphics[width=5in]{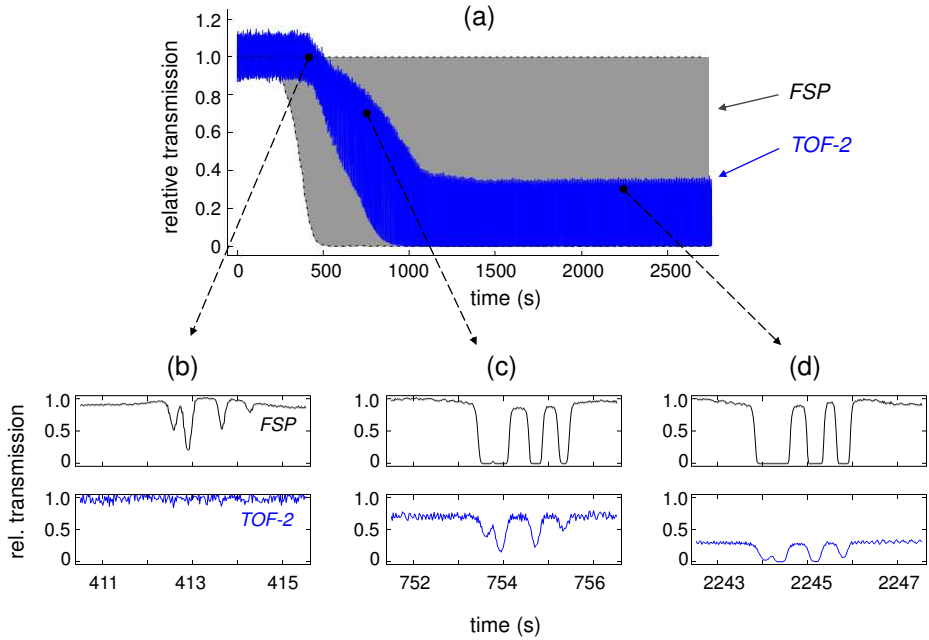}
\caption{Analogous run to Fig. \protect\ref{fig:TOF1}, but with the TOF heater turned on at a nominal value of 200$^{o}$C (much higher than the vacuum chamber temperature of 60$^{o}$C). In this case, the TOF (TOF-2) transmission degradation is much less severe, with a final TOF-2 transmission of 30\% of its initial value. Comparing the TOF-2 dip depths in part(d) with the TOF-1 dips in Fig. \protect\ref{fig:TOF1}(d) suggests that the TOF heating unit can enable relatively high transmission while maintaining sufficient Rb density.}
\label{fig:TOF2}
\end{figure}

\begin{figure}[t]
\centering\includegraphics[width=5in]{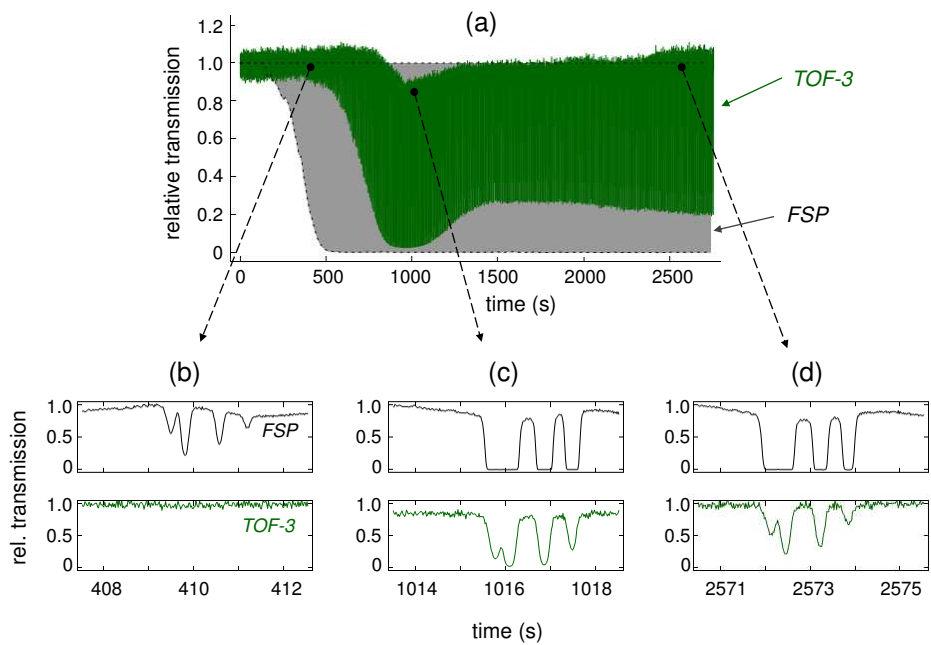}
\caption{Detailed results of a third run performed with lower Rb density, and a higher TOF heating unit temperature (265$^{o}$C). These results show the ability to preserve 100\% TOF transmission (TOF-3) in relatively low Rb density. Comparing the TOF-3 data in parts (c) and (d) suggests the possibility of preserving high transmission in higher Rb densities by using an even hotter TOF heating unit.}
\label{fig:TOF3}
\end{figure}

The behavior of the TOF system was studied by continuously sweeping the laser frequency back and forth across the same 20 GHz range (red sawtooth in Fig. \ref{fig:Psat}(a)) while Rb vapor density was introduced into the vacuum chamber.  Monitoring the absorption dips and overall transmission of the TOF signal as a function of time provided a clear picture of transmission degradation and preservation during a run. The results of three key runs are shown in Figures \ref{fig:TOF1} - \ref{fig:TOF3}. For all of these runs, the TOF power was kept at $< 1$ nW to avoid any complications due to saturation effects.

Fig. \ref{fig:TOF1}(a) shows the results of a run with the TOF heating unit intentionally turned off. The gray trace in the background shows the transmission of the free-space probe (FSP) beam passing through the chamber; this had a long optical path length of 12 cm and was therefore very sensitive to the introduction of Rb vapor at the beginning of a run. The full 2700 second run corresponds to 540 consecutive sets of absorption dips (five second sweep), so the displayed FSP data appears as a thick gray shaded region.  The thick red curve in the foreground is the TOF signal recorded at the same time. The overall TOF transmission rapidly degrades roughly 600 seconds into the run due to Rb accumulation on the TOF surface.

Figures \ref{fig:TOF1}(b)-(d) show zoom-ins of individual five-second sweeps at three points during the run. At 413 seconds (part (b)) significant Rb vapor has entered the system; the sensitive FSP begins to show absorption dips, while dips are not yet apparent in the TOF (which has a much shorter effective interaction length on the order of 1 cm).  By 593 seconds (part (c)) the Rb density has substantially increased; the TOF is showing deep absorption dips, but its overall transmission has already degraded to 60\% of its initial value. At 2246 seconds (part (d)), the overall TOF transmission has dropped to 1.5\% of its initial value. Remarkably, the remaining TOF signal still shows the Rb absorption dips.

In comparison, Figure \ref{fig:TOF2} shows the results of a second run with the TOF heating unit turned on at a nominal value of 200$^{o}$C (measured by the internal tc of the ceramic heater). The idea was to raise the TOF surface temperature higher than the chamber (60$^{o}$C). In Fig. \ref{fig:TOF2}(a), transmission degradation is seen to occur on a similar time-scale as Fig. \ref{fig:TOF1}(a), but with substantially less overall loss and a final TOF transmission at 30\% of its initial value. The zoom-ins in Figs. \ref{fig:TOF2}(b)-(d) show that the heating unit enables the preservation of relatively high steady-state TOF transmission (30\%) in a relatively high Rb vapor density, as indicated by the nearly 100\% TOF dips in part (d).

The primary ``heater off'' vs. ``heater on'' summary data in Fig. \ref{fig:mainresults} was extracted from these two data sets (Figs. \ref{fig:TOF1}(a) vs. \ref{fig:TOF2}(a)). The solid and dashed black lines in Fig. \ref{fig:mainresults} track the FSP beam while on resonance with the strong ``dip 2'' transition, while the red and blue curves correspond to TOF transmission data extracted away from any Rb resonance. This off-resonant TOF data provides the best indication of the overall TOF transmission vs. time for these runs.

Experimentally ensuring approximately the same Rb density vs. time profiles enabled a fair comparison between these ``heater off'' vs. ``heater on'' runs. This was confirmed using the depth of the FSP absorption dips to measure the rapidly changing relative density in the chamber at the beginning of a run; and the depths of the TOF dips once they became apparent after  $t \sim 500$ seconds. Because of the difficulties in accurately fitting the absorption profiles of 100\% dips \cite{siddons08}, we simply verified that the optical depths (OD's) of the smallest TOF dip (dip 4) were roughly the same for the latter parts of the runs. The OD's of this particular signal raised to maximum values near 2.0 at $t \sim 1200$ seconds, and then slowly dropped towards values near 1.5 at the end of each run.

Fig. \ref{fig:TOF3} shows the results of a third run which is not directly comparable to the first two runs. For convenience, the same Rb heating procedure was used, but the vacuum system was not thoroughly baked-out prior to the run. Consequently, the Rb vapor density was substantially lower during the run. In addition, the temperature of the TOF heating unit was raised from the nominal value of 200$^{o}$C to 265$^{o}$C. As shown in Fig. \ref{fig:TOF3}(a), this combination of lower density and higher TOF temperature allowed preservation of 100\% transmission through the TOF. For this run, the OD (of the TOF dip-4 signal) only raised to maximum value of roughly 1.0 at $t \sim 900$ seconds, and then dropped to a steady-state value roughly 0.5 in the second half of the run. The zoom-ins in Figs. \ref{fig:TOF3}(b)-(d) show the corresponding increase and decrease of TOF dip depths. Although not conclusive, these results suggest the possibility of achieving 100\% TOF transmission in even higher Rb densities by further TOF heating.

\section{TOF failure mechanisms}
\label{sec:TOFfailures}

\begin{figure}[t]
\centering\includegraphics[width=5in]{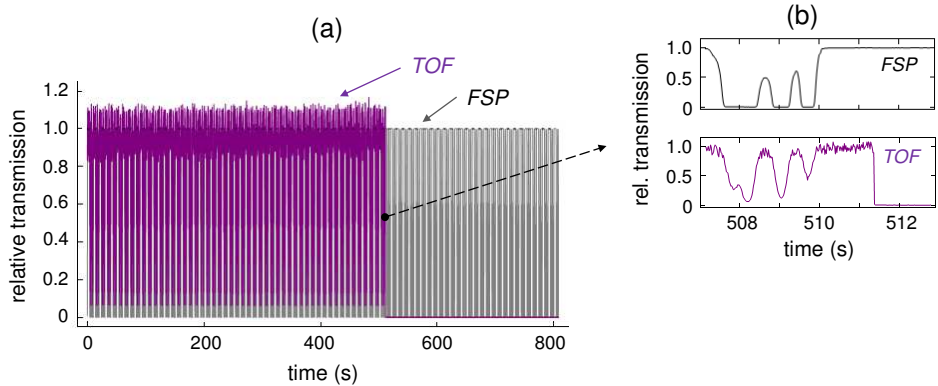}
\caption{(a) Data showing an otherwise successful run in which the TOF transmission (purple trace) suddenly drops to zero at $t= 511$ seconds. (b) A zoom-in shows that the complete transmission drop occurs on time-scale of much less than 1 second, suggestive of a ``fracture'' in the TOF. The TOF data is normalized to 1 to highlight the transmission drop. These sudden transmission drops were the primary failure mechanism that prevented us from indefinitely using our heated-TOF's in Rb vapor.}
\label{fig:instantbreak}
\end{figure}

The ability to indefinitely maintain relatively high overall TOF transmission in a standard ``heater on'' run (such as Fig. \ref{fig:TOF2}) was primarily limited by a failure mechanism in which the TOF transmission would instantaneously drop to zero. Fig. \ref{fig:instantbreak} shows the data capturing one of these TOF failures. In analogy with Figs. \ref{fig:TOF1}-\ref{fig:TOF3}, the TOF data (purple trace) is plotted on top of the FSP data (grey trace).  At $t = 511$ seconds into an otherwise stable run  (part (a)), the TOF transmission suddenly drops to zero; the zoom-in (part(b)) shows that the time-scale of the drop is much less than 1 second.

This rapid time-scale is suggestive of some type of ``fracture'' within the TOF itself. Subsequent examination of the TOF's with visible red-light usually showed large scattering out in the tapered region. The idea of a fracture is further supported by the fact that a smaller number of the failed runs resulted in the TOF being physically broken into two pieces. We note that these ``fractures'' were also observed while maintaining a TOF under vacuum without Rb vapor.

It is possible, however, that these rapid transmission drops were due to large contaminant particles ``landing'' on the TOF surface. Fujiwara et.al. have recently shown rapid transmission drops (albeit not to zero) due to dust particles \cite{fujiwara12}, and similar effects have been seen by placing polystyrene clusters on TOF's \cite{gregor09}. 

In contrast, transmission degradation due to Rb accumulation in our system (summarized in Fig. \ref{fig:mainresults}) occurred on a time-scale of several minutes. We note that similar time-scale transmission degradation was also observed during initial vacuum system bake-outs with the TOF heating unit turned off; presumably water vapor and other contaminants accumulating on the TOF surface degraded transmission in a similar manner. From a practical point of view, keeping the TOF heating unit turned on at 200$^{o}$C entirely eliminated transmission degradation during bake-outs. 

Similar slow time-scale transmission degradation was also observed when we attempted to raise the TOF heating unit temperature above 265$^{o}$C. We suspect outgassing of the epoxy (and/or other materials) at these higher temperatures contaminated the TOF surfaces. This effect prevented us from using higher TOF temperatures in an attempt to demonstrate preservation of TOF transmission at higher Rb densities.

\section{Summary and conclusions}
\label{sec:summary}

Transmission loss due to Rb accumulation on the surface of a TOF is currently one of the main limiting factors in the utility of these systems for low-power nonlinear optics \cite{spillane08,hendrickson10,salit11}. In this paper we have described real-time observations of TOF transmission degradation as the density of the surrounding Rb vapor was increased.  By comparing the case of non-heated TOF's with heated TOF's, we showed that a heating unit can be effectively used to preserve relatively high TOF transmission, while maintaining the relatively high atomic densities needed for many low-power nonlinear optics applications. We achieved steady-state high TOF transmission conditions (like Figs. \ref{fig:TOF2} and \ref{fig:TOF3}) that lasted from several hours to several days in some cases, with suspected spontaneous ``fractures'' of the TOF (like Fig. \ref{fig:instantbreak}) being the primary failure mechanism.

The interaction of Rb with silica surfaces is a complex process \cite{ma09}. In analogy to heating the windows in Rb vapor cell experiments, the basic idea of raising the TOF surface temperature higher than the surrounding environment is simply to prevent the initial accumulation of Rb on the tapered waist region. Although not as successful, we also found that our TOF heater could be used to remove Rb once accumulated, resulting in a partial recovery of TOF transmission. The preservation and recovery of TOF transmission has also been observed using a (relatively) high-power internally guided laser beam \cite{hendrickson09}, and the possibility of external LIAD techniques with TOF systems may also be useful \cite{slepkov08}. The results presented here are encouraging for the advancement of the basic ``TOF in Rb system'' towards practical low-power nonlinear optical devices.

\section*{Acknowledgments}
This work was supported in part by the DARPA ZOE program (Contract No. W31P4Q-09-C-0566).  The authors thank D. Dubbel for assistance with the design and fabrication of the TOF heating system, and acknowledge fruitful discussions with S.M Hendrickson and T. Gougousi.

\end{document}